\documentclass[12pt,preprint]{aastex}
\begin{document}

\title{SDSSJ103913.70+533029.7: A Super Star Cluster in the Outskirts of a 
Galaxy Merger}

\author{
Gillian R. Knapp\altaffilmark{1},
Christy A. Tremonti\altaffilmark{2},
Constance M.  Rockosi\altaffilmark{3},
David J. Schlegel\altaffilmark{4},
Brian Yanny\altaffilmark{5},
Timothy C. Beers\altaffilmark{6},
Carlos Allende Prieto\altaffilmark{7},
Ron Wilhelm\altaffilmark{8},
Robert H. Lupton\altaffilmark{1},
James E. Gunn\altaffilmark{1},
Martin Niederste-Ostholt\altaffilmark{1},
Donald P. Schneider\altaffilmark{9},
Kevin Covey\altaffilmark{10},
Anil Seth\altaffilmark{10},
\v{Z}eljko Ivezi\'c\altaffilmark{10},
Daniel J. Eisenstein\altaffilmark{2},
Joe Helmboldt\altaffilmark{11},
Douglas P. Finkbeiner\altaffilmark{1},
Nikhil Padmanabhan\altaffilmark{12},
Scot J. Kleinman\altaffilmark{13},
Dan Long\altaffilmark{13},
Stephanie A. Snedden\altaffilmark{13},
Atsuko Nitta\altaffilmark{13},
Michael Harvanek\altaffilmark{13},
Jurek Krzesinski\altaffilmark{13,14},
Howard J. Brewington\altaffilmark{13},
John C. Barentine\altaffilmark{13},
Peter R. Newman\altaffilmark{13},
Eric H. Nielsen Jr.\altaffilmark{5},
Masataka Fukugita\altaffilmark{15},
J. Brinkmann\altaffilmark{13}
}
\altaffiltext{1}{Department of Astrophysical Sciences, Princeton University, 
Princeton, NJ 08544
\label{Princeton}}
\altaffiltext{2}{Department of Astronomy and Steward Observatory,
University of Arizona, 933 N. Cherry Avenue, Tucson, AZ 85721-0065
\label{Arizona}}
\altaffiltext{3}{Lick Observatory, University of California, Santa Cruz,
CA 95064
\label{UCSC}}
\altaffiltext{4}{Lawrence Berkeley Laboratory, 1 Cyclotron Road, Berkeley,
CA 94720
\label{LBL}}
\altaffiltext{5}{Fermi National Accelerator Laboratory, P.O. 
Box 500, Batavia, IL 60510
\label{Fermilab}}
\altaffiltext{6}{Department of Physics and Astronomy and JINA: Joint
Institute for Nuclear Astrophysics, Michigan
State University, East Lansing, MI 48824
\label{Michigan}}
\altaffiltext{7}{McDonald Observatory and
Department of Astronomy, University of Texas at Austin,
Austin, TX 78712
\label{UT}}
\altaffiltext{8}{Department of Physics, Texas Tech University, Lubbock, 
TX 79409
\label{Texastech}}
\altaffiltext{9}{Department of Astronomy and Astrophysics,
The Pennsylvania State University,
University Park, PA 16802
\label{PennState}}
\altaffiltext{10}{University of Washington, Department of 
Astronomy, Box 351580, Seattle, WA 98195
\label{Washington}}
\altaffiltext{11}{Department of Astronomy, New Mexico State University,
Box 30001, Dept. 4500, Las Cruces, NM 88003-8001
\label{NMSU}}
\altaffiltext{12}{Department of Physics, Joseph Henry Laboratories, Princeton
University, Princeton, NJ 08544
\label{Princetonphys}}
\altaffiltext{13}{Apache Point Observatory, PO Box 59, Sunspot,
NM 88349
\label{APO}}
\altaffiltext{14}{Mt. Suhora Observatory, Cracow Pedagogical Observatory,
ul. Podchorazych 2, 30-084, Cracow, Poland
\label{Cracow}}
\altaffiltext{15}{Institute for Cosmic Ray Research, University of
Tokyo, 5-1-5 Kashiwa, Kashiwa City, Chiba 277-8582, Japan
\label{ICRR}}
%
%------------------------------------------------------------------------------
% ABSTRACT
%------------------------------------------------------------------------------

\begin{abstract}
We describe the serendipitous discovery in the spectroscopic data of the
Sloan Digital Sky Survey of a star-like object, SDSSJ103913.70+533029.7, at a
heliocentric radial velocity of +1012 $\rm km~s^{-1}$.  Its proximity
in position and velocity to the spiral galaxy NGC 3310 suggests
an association with the galaxy. At this distance, SDSSJ103913.70+533029.7
has the luminosity of a super star cluster and a projected distance of
17 kpc from NGC 3310.  Its spectroscopic and photometric properties
imply a mass of $\rm > 10^6~ M_{\odot}$ and an age close to that of
the tidal shells seen around NGC 3310, suggesting that it formed in
the event which formed the shells.

\end{abstract}
\keywords{galaxies:star clusters--galaxies: individual(NGC
3310)--galaxies:starburst--galaxies:formation}

%------------------------------------------------------------------------------
% INTRODUCTION
%------------------------------------------------------------------------------
\section{INTRODUCTION}
\label{sec_intro}

Super star clusters (van den Bergh 1971;
Arp \& Sandage 1985; Schweizer 1986; Lan\c{c}on \& Boily 2000;
Lamers, Smith \& Nota 2004; Gallagher \& Smith 2004)
are compact (diameters less than 20
pc) high luminosity ($\rm M_V~<$ -10) clusters whose properties
resemble those of globular clusters 
except that they are young, often much less than 1 Gyr in age. As
such, they have importance for our understanding of the formation
of globular cluster systems and by extension of the galaxies of 
which they are members.

Large numbers of these systems have been discovered and characterized
in the last 15 years or so, especially from high angular resolution
imaging with {\it Hubble Space Telescope} and spectroscopic
observations with new very large ground-based optical telescopes.
While super star clusters have not yet been seen in the Milky Way or in M31, they
have been discovered in large numbers in several environments: in starbursting
gas--rich dwarf irregular galaxies (Melnick, Moles \& Terlevich 1985, O'Connell,
Gallagher \& Hunter 1994, Ho \& Filippenko 1996a,b, Gelatt, Hunter
\& Gallagher 2001, Billett et al. 2002, Vanzi 2003), in 
more massive starburst galaxies
(Holtzmann et al. 1992, Whitmore et al. 1993, Meurer et al. 1995,
Whitmore \& Schweitzer 1995, Zepf et al. 1999, Tremonti et al. 2001,
Lipscy \& Plavchan 2004, Turner \& Beck 2004, McCrady et al. 2005, Melo
et al. 2005), many of which show signs of recent mergers or
interactions, and in the compact group Stephan's quintet
(Gallagher et al. 2001). In some of these galaxies, such as M82, 
NGC 4038/9, NGC 1275, etc., super star clusters
incorporate a significant fraction of the starburst (Lipscy \& Plavchan 2004,
McCrady et al. 2005) and are preferentially found in the inner regions of
the galaxy (Whitmore \& Schweizer 1995, Meurer et al. 1995). Their masses 
are measured to be up to $\rm 10^6 ~ M_{\odot}$ (McCrady et al. 2005),
and their metalicities range from [Fe/H] = -1.5 in dwarf irregular galaxies
to several times solar in the large galaxy mergers.

NGC 3310 is a small spiral galaxy ($\rm M_B$ = -19.6; D = 
14 Mpc, adopting the Hubble distance found using $\rm H_{\circ}
~ = ~ 72~km~s^{-1}~Mpc^{-1}$)
which has a disturbed morphology both in its optical light and  HI
distribution (Kregel \& Sancisi 2001; Mulder \& van Driel 1996; de 
Grijs et al. 2003a,b; Wehner \& Gallagher 2005), and whose inner regions
underwent a burst of star formation about 30 Myr ago, with the formation of
several hundred super star clusters (de Grijs et al. 2003b). The present
paper describes the 
serendipitous discovery 
of SDSSJ103913.70+533029.7 (hereafter SDSS1039+53), an unresolved
($< 1''$)
object with a radial velocity of about 1000 $\rm km~s^{-1}$.
Several lines of evidence suggest that it is associated with NGC 3310. If
so, its luminosity, colors and spectrum point to its being a compact, 
high-luminosity super star cluster at a projected distance of 17 kpc
from NGC 3310. By contrast, all previously-known
super star clusters lie in the inner kpc
or two of their parent galaxy. 

The relevant details of the SDSS observations are summarized in the 
next section. Section 3 describes the properties and nature of SDSS1039+53. We
discuss the possibility that this object is a high-velocity Galactic star,
but favor the evidence associating it with NGC 3310. Under this assumption,
its properties are shown to be consistent with its being a super star
cluster of age similar to that of the recent merger/interaction involving
NGC 3310.  The SDSS imaging also confirms the presence of stellar shells
around NGC 3310 (Wehner \& Gallagher 2005) and shows the
disturbed morphology of the galaxy.  The
discussion and conclusions are given in Section 4.
%------------------------------------------------------------------------------
% SDSS
%------------------------------------------------------------------------------
\section{The Sloan Digital Sky Survey}

The Sloan Digital Sky Survey (SDSS) is a 5-band photometric survey of
about 10,000 square degrees of the Northern sky to a depth of about
22.5 ($r$ magnitude, point source) and a concurrent
redshift survey of up to a million galaxies and 100,000 quasars
selected from the imaging survey (York et al. 2000). 
The SDSS camera (Gunn et al. 1998) mounted on a dedicated 2.5 meter
telescope (Gunn et al. 2005)
at Apache Point Observatory (APO), New Mexico, acquires
imaging data in five bands,
$u$, $g$, $r$, $i$, and $z$, centered at approximate effective wavelengths
of 3551, 4686, 6166, 7480 and 8932 \AA{} (Fukugita et al. 1996).
The imaging data are automatically reduced through a series of software
pipelines (Lupton et al. 2001, 2003;
Pier et al. 2003; Ivezi\'c et al. 2004).
The instrumental fluxes are calibrated via a network of primary and
secondary stellar flux standards to $\rm AB_{\nu}$ magnitudes (Oke \&
Gunn 1983; Fukugita et al. 1996; Hogg et al. 2001; Smith et al. 2002,
Tucker et al. 2005),
and the absolute positions are calibrated using standard astrometric 
catalogues (Pier et al. 2003).

Targets for spectroscopy
are selected from the imaging data on the basis 
of their photometric properties. As well as the primary SDSS targets
(galaxies and quasars),
stars in many different locations of color-magnitude space are selected
to provide backup
targets in regions of low galaxy density and serve as 
spectrophotometric standards. The target objects are mapped
(Blanton et al. 2003) onto $3^{\circ}$ diameter aluminum fiber plug
plates which feed the spectrographs.  The pair of dual fiber-fed 
spectrographs (Uomoto et al. 1999) can observe 640 spectra at one 
time with a wavelength coverage of 3800 - 9200 \AA{} and a resolution 
of 1800 to 2100. 
The spectra are extracted from CCD images, calibrated, and 
corrected for sky emission and absorption. The
resulting calibrated spectra are fit to a series of
templates of galaxies, quasars and stars to derive the spectral
classification, redshift and redshift error of each object
(D. Schlegel, in preparation: see {\bf http://spectro.princeton.edu}). 
Stellar templates
are taken directly from SDSS spectra and calibrated with respect
to spectral type and radial velocity using the ELODIE stellar
library ({Prugniel \& Soubiran 2001; Moultaka et al. 2004).
The SDSS data are described in the data release papers 
by Abazajian et al. (2003, 2004, 2005) and Adelman-McCarthy
et al. (2006)
and documented at web sites listed therein and at 
{\bf http://www.sdss.org}.

We examined the spectroscopic data in the SDSS archives
as of January 12, 2005, selecting objects spectroscopically classified as stars.
The sample 
was further selected by magnitude: $g < $ 20, and by 
color: 0.8 $< ~ u-g ~ <$ 1.5, $g-r ~<$ 0.5, roughly the colors
of halo turnoff stars (F subdwarfs) and horizontal branch stars
(cf. Yanny et al. 2000, Newberg et al. 2002, Sirko 
et al. 2004a). During the selection, magnitudes were not
corrected for interstellar extinction. 
SDSS has obtained spectra of many tens of thousands of stars with these
properties; in all, 40746 spectra were
found in the data base that satisfied these selection criteria.
The heliocentric radial velocities of 
the stars were corrected
to the Galactic standard of rest assuming a solar velocity of 16.6 
$\rm km~s^{-1}$ towards $\rm \alpha(2000)$ = $\rm 17^h~49^m~58.7^s$,
$\rm \delta(2000)$ = $\rm +28^{\circ}~07'~04''$,
and a motion of the LSR of 220 $\rm km~s^{-1}$ towards 
$\rm \alpha(2000)$ = $\rm 21^h~12^m~01.1^s$, $\rm \delta(2000)$
= $\rm +48^{\circ}~19'~47''$.
The distribution of velocities for this large sample is  Gaussian
with a standard deviation of 102 $\rm km~s^{-1}$ (Sirko et al. 2004b);
however, one object, SDSS1039+53, has 
a Galactocentric velocity well outside this range,
+1067$\pm$ 19 $\rm km~s^{-1}$,  greatly in excess of the
expected escape velocity from the Galaxy. Various properties of 
SDSS1039+53 are summarized in Table 1. Its SDSS spectrum is
shown in Figure 1, and its colors, corrected
for Galactic extinction using the values given by Schlegel,
Finkbeiner \& Davis (1998), are compared with those of a sample of
stars observed by SDSS from Finlator et al. (2000) in Figure 2.

\section{SDSS1039+53: Galactic or Extragalactic?}

SDSS1039+53 was observed in two imaging runs.  In both observations,
the object is classified as a point source in all five bands,
except for the $u$ band image taken in poorer seeing (run 2735). In the
run with better
seeing (run 2821) the object is unresolved at the PSF diameter,
$\rm 0.9''$, in the $r$ and $i$ bands.
Its colors (Figure 2) are close to but not identical with
those of halo/thick disk F/G stars.  
The object has no detectable proper motion; USNO-B + SDSS gives
$\rm \mu_{\alpha}$ = 1 $\pm$ 3.9 mas~$\rm yr^{-1}$,
$\rm \mu_{\delta}$
 = $-$2 $\pm$ 3.9 mas~$\rm yr^{-1}$ (Munn et al. 2004).

There is only one spectrum of SDSS1039+53 in the
SDSS data archives, that shown in Figure 1, and it is similar to
that of a middle-F star.  Thus at first glance SDSS1039+53 appears
to be a star with a velocity well outside the velocity range
of Galactic stars, and is
the only such object so far discovered in the SDSS
spectroscopic data base. Given the extraordinary value of the radial
velocity, we performed a number of tests to investigate the reality
of this measurement.
First, the spectrum was analyzed using two different 
codes and template sets. Cross-correlation analysis (SubbaRao
et al. 2002)
gives a heliocentric velocity of 1028 $\pm$ 23 $\rm km~s^{-1}$, while
a direct $\chi^2$ minimization fit (D.J. Schlegel, unpublished)
gives 1012 $\pm$ 19 $\rm km~s^{-1}$.
The use of different templates, with types
from late B to
F, produces  a scatter of about 10 $\rm km~s^{-1}$, and fitting a 
polynomial to the centers of the Balmer lines gives 1009 $\pm$
9 $\rm km~s^{-1}$.
Second, the spectrum is actually the composite of two spectra, measured by the
red and blue spectrographs. Both yield the same radial velocity.
Third, there are five stars on plate 1010 which are also observed
on other plates, and the measured radial
velocities agree to within the errors.
In total, there are 30
stars in the selected color and magnitude range
observed in plate 1010 in addition to SDSS1039+53,
and the measured velocities span the range -140 $\rm km~s^{-1}$
to +213 $\rm km~s^{-1}$. Fourth, the sky spectrum was examined.
Figure 1 shows the 
sky spectrum in the
region 5400 to 6000 \AA{}, where very strong night-sky emission
from the Na D lines and the [OI] $\lambda$5577 airglow lie.
Were there an error in the wavelength calibration of the
spectrum of SDSS1039+53, the night sky subtraction would
show very large residuals in this wavelength region, when in
fact there is no discernable residual at NaD and only a small 
residual at 5577 \AA{}. 

A final, and definitive test was supplied by a spectrum obtained
with the Marcario Low Resolution Spectrograph (Hill et al. 1998)
on the Hobby-Eberly telescope on 14 February 2005.  This spectrum, 
which covered the H$\alpha$ line at a resolution of approximately
1000, yields a heliocentric radial velocity of 1033 $\pm$ 70
$\rm km~s^{-1}$, consistent with the results from the SDSS data.

Therefore, SDSS1039+53 could be a Galactic 
high-velocity star, like that recently discovered by Brown et al.
(2005). Several methods were used to derive the stellar parameters
under this assumption.
Fitting the flux--calibrated spectrum to synthetic spectra based on
Kurucz (1979, 1993) model atmospheres using the methods
of Allende Prieto et al. (2005) gives: 
T$_{\rm eff}$ =
6237 $\pm 247$~K, log g = 2.00$ \pm 0.26$, [Fe/H] $= -0.37 \pm 0.17$, and
badly overestimates the strength of the Ca II
K line. The techniques described by Wilhelm et al. (1999) yield: T$_{\rm eff}$
= 6456 K, log g $= 2.27$, and [Fe/H] $= -2.44$, with the abundance
derived primarily from the CaII K line. Other metal lines give a  much higher
value, [Fe/H] = $-0.71$. The spectrum and colors are consistent with
classification of SDSS1039+53 as a horizontal-branch star at 60
kpc. However, the line widths for H$\epsilon$ and
H$\delta$ are much larger than that of H$\gamma$. Finally, the abundance
obtained using the Beers et al. (1999) calibration of the CaII K
line and the colors is [Fe/H] =
$-1.88$. Taken as a whole, these results suggest that the spectrum of this
object is composite, which might be expected if it is not a single star, but
the integrated light from a cluster of stars. 

If, on the other hand,
SDSS1039+53 is extragalactic, as suggested by its radial velocity,
its very compact light distribution (with an inferred
diameter of $\leq$ 40 pc) suggests two possible interpretations:
it may be a compact dwarf galaxy, like those formed by tidal 
disruption in clusters (Drinkwater et al. 2003, Ha\c{s}egan et al. 2005), or
a star cluster like those seen in starburst galaxies. SDSS1039+53 does not
lie in a cluster of galaxies, but is 
about $\rm 4.1'$ from the Sbc galaxy NGC 3310, which has a very similar radial
velocity,
$\rm V_{helio}$ = 993 $\rm km~s^{-1}$.  Further, NGC 3310 itself is 
a galaxy which has recently undergone a merger with one or more
gas-rich systems: it has extensive HI tidal tails and is 
surrounded by a low surface brightness shell system (Kregel \&
Sancisi 2001; Wehner \& Gallagher 2005). Thus the more likely
interpretation is that SDSS1039+53 is a compact star cluster associated with
NGC 3310. 

\section{Discussion and Conclusions}

NGC 3310 is at a distance of about 14 Mpc and is a ``minor merger'', with
extended HI tidal tails (Kregel \& Sancisi 2001), stellar shells surrounding
the main disk (Wehner \& Gallagher 2005) and a large-scale
starburst in its inner regions (de Grijs et al. 2003b) which is 
considerably younger than the inferred age of the merger (see
below).  The SDSS image
of this galaxy and its surroundings is shown in Figure 3. Tidal
shells can be seen, as well as the actively star-forming spiral arms
which have a distorted and chaotic morphology.  The image also shows
SDSS1039+53, at a distance of 4.14 arcminutes from the center of
NGC 3310.  The coincidence in velocity and sky position strongly
suggests that these two unusual objects are associated.  If so, the 
absolute magnitude of SDSS1039+53 is M$_r$ = -11.3 ($\rm
M_V ~ \sim$ -11.2, using the conversions given by Fukugita et al.
1996).  This luminosity is much higher than
those of the Galactic globular clusters
(cf. Peterson 1993 and Djorgovski 1993)
but is similar to those of super star clusters.

What is remarkable about SDSS1039+53 is its
distance from the galaxy
nucleus (4.14 arcminutes, 17~kpc). Meurer et al. (1995) note that
super star clusters are typically found ``at the very heart of
starbursts''. In their ultraviolet imaging survey, over 90\%
of super star clusters were found where the local surface brightness
was within 1.5 magnitudes $\rm arcsec^{-2}$ of its peak value. SDSS1039+53, 
in contrast, is well outside the optical extent of NGC 3310 as seen in 
Figure~3, and considerably beyond
the de Vaucouleurs radius ($\rm R_{25}$ = 1.55 arcmin, de Vaucouleurs et 
al. 1991).  The cluster's projected distance is equivalent to
$\sim$20 disk scale lengths ($\rm 
r_{\circ}(I)$ = 12.4 arcsec = 0.84 kpc, S\'anchez-Portal et al.
2004). This would be untypical -- but not unheard of -- for a 
globular cluster; roughly 95\% of the Milky Way's globular
clusters are found within $\sim$10$r_0$ 
($r_0$ = 2.8~kpc, Ohja 2001), with the most distant globular
cluster lying at 42~$r_0$, 117 kpc (Harris 1996).

To determine an age for the star cluster we fit Bruzual \& Charlot
(2003) simple stellar population models (SSPs) to both the optical
spectrum and to the SDSS $ugriz$ colors (both the spectrum and 
colors were corrected for Galactic foreground reddening and
extinction using the maps of Schlegel, Finkbeiner \& Davis 1998).
Because the cluster lies so far outside the optical and HI disk of 
NGC~3310, we assumed no intrinsic reddening.  We adopted a 
metalicity of Z = 0.008 (60\% solar) since this is in good accord
with the nebular abundances in HII regions outside the nucleus
(Pastoriza et al. 1993) and with metalicities derived for star
clusters from {\it HST} multiband photometry (de Grijs et al.
2003b).  We compared the data to a grid of 110 models with ages
ranging from 10~Myr to 1~Gyr. The $\chi^2$ evolution of the models 
along with the best fits are shown in Figure~4.  The optical spectrum
is fit over the full wavelength range available, 3800 - 9200 \AA{}.
The best-fit spectrum has an age of 570~Myr. The photometry provides
additional information, because the SDSS spectra do not cover the 
$u$ and $z$ bands; the photometry alone fits best to an SSP model
of age 720~Myr. These two independent fits provide some measure
for the systematic uncertainties in the SDSS photometry and 
spectroscopy, and give a cluster age of 700$\pm$150~Myr. The
model cluster mass is 1.4$\rm \times 10^6 ~ M_{\odot}$,
assuming a Chabrier (2003) Galactic Initial Mass Function
(a Salpeter IMF extending down to 0.1~$\rm M_{\odot}$ would increase
the mass estimate by a factor of 1.4).

The derived age and mass of SDSS1039+53 place it squarely in the 
r\'egime of super star clusters (age $\lesssim$ 1~Gyr, mass = 
$\rm 10^5$ - $\rm 10^8 ~ M_{\odot}$, O'Connell 2004). However,
when compared to the inner disk cluster population in NGC~3310, 
SDSS1039+53 stands out as being both relatively old and massive.
de Grijs et al. (2003a,b) characterized the cluster population
in the inner 33 arcseconds of NGC~3310 using multi-band {\it
HST} photometry. Out of 174 clusters, they find only a few percent
with $\rm M ~ \sim ~ 10^6 ~ M_{\odot}$ and roughly 10\% with
ages $>$ 100 Myr.  Only one cluster satisfies both criteria.
The mean age of the cluster population is around 30 Myr, and
de Grijs et al. suggest a burst duration of 40 Myr. Thus 
SDSS1039+53 is considerably older than the most recent starburst
event in the galaxy disk.

NGC 3310 is surrounded by shells and tidal debris (see Figure 3)
and recent observations have found additional features
extending to distances of about 15 kpc
from the galaxy.  Kregel \& Sancisi (2001) observed HI tails to the
north and south of the galaxy, while Wehner \& Gallagher (2005) find a 
faint stellar arc to the north and east. These observations suggest an age 
for the encounter which produced the faint stellar arc of
about 0.5 Gyr, close to the age inferred for SDSS1039+53. The
position of SDSS1039+53 lies close to that of the faint outer
stellar arc - near feature ``N'' in Figures 2 and 4 of Wehner and
Gallagher (2005). (This feature has a surface brightness of
$\rm \mu_V$ = 22.55~mag~$\rm arcsec^2$ and is not visible in the
shallower SDSS images).  Wehner \& Gallagher (2005) suggest
that the outer stellar arc is tidal debris from a disrupted
dwarf companion. However, our finding of a comparatively young
cluster in this location suggests that the faint light could be
composed of stars formed in the merger event. Obtaining accurate
colors for the faint light should help clarify this situation.
Although it is not necessary to invoke formation of
SDSS0139+53 at this location -- had it formed in the center 
of the galaxy it would have needed an average velocity in the
plane of the sky with respect to the
galaxy of only 35 $\rm km~s^{-1}$ to reach its present location --
its properties identify it with the merger event which
took place in NGC 3310 and identify it as a young globular cluster
associated with and formed from the far-flung d\'ebris of the merger.

While most of the super star clusters found to date in starburst dwarfs and
merging galaxies are in the inner regions of the galaxies, it is
also the case that for almost none of these galaxies have the
surroundings been searched.  Yet star formation has been observed well
outside the optical disks in a small number of galaxies. For example, Hibbard
et al. (2005) identify star formation in the tidal tails of the
Antennae (NGC 4038/9) in GALEX images; small HII regions have been
discovered at distances as large as 30 kpc from their galaxy
(Ryan-Weber et al. 2004), and in most of the cases discussed in that
paper there is evidence for tidal
disturbance of the associated galaxy; and HST imaging
shows star clusters and dwarf galaxies in the tidal tails of merger
galaxies which are themselves undergoing intense star formation
(Knierman et al. 2003; Tran et al. 2003; Bastian et al. 2005).   Thus
it is possible that there are many more super star clusters
at large distances from NGC 3310 -- the field (Figure
3) contains several more sources of bluish color -- and that
young globular cluster systems can be identified around this and other
galaxies.

Finally, we compare SDSS1039+53 with SDSS090745.0+024507,
the hyper-velocity star discovered by Brown et al. (2005).  As Figure
2 shows, SDSS0907+02 is much bluer than SDSS1039+53, with
the colors of a young star.  Further, there are no plausible host
galaxies within many degrees. Thus unlike
SDSS1039+53, the observational data for SDSS0907+02 strongly support
the interpretation of Brown et al. (2005): that it is a 
star escaping the Galaxy.  So far, SDSS spectroscopy has not revealed
any more hyper-velocity Galactic stars.
%
%----------------------------------------------------------------------
%  FIGURE 1  Spectrum
%----------------------------------------------------------------------
\begin{figure}[tb]
\figurenum{1}
\epsscale{0.9}
\plotone{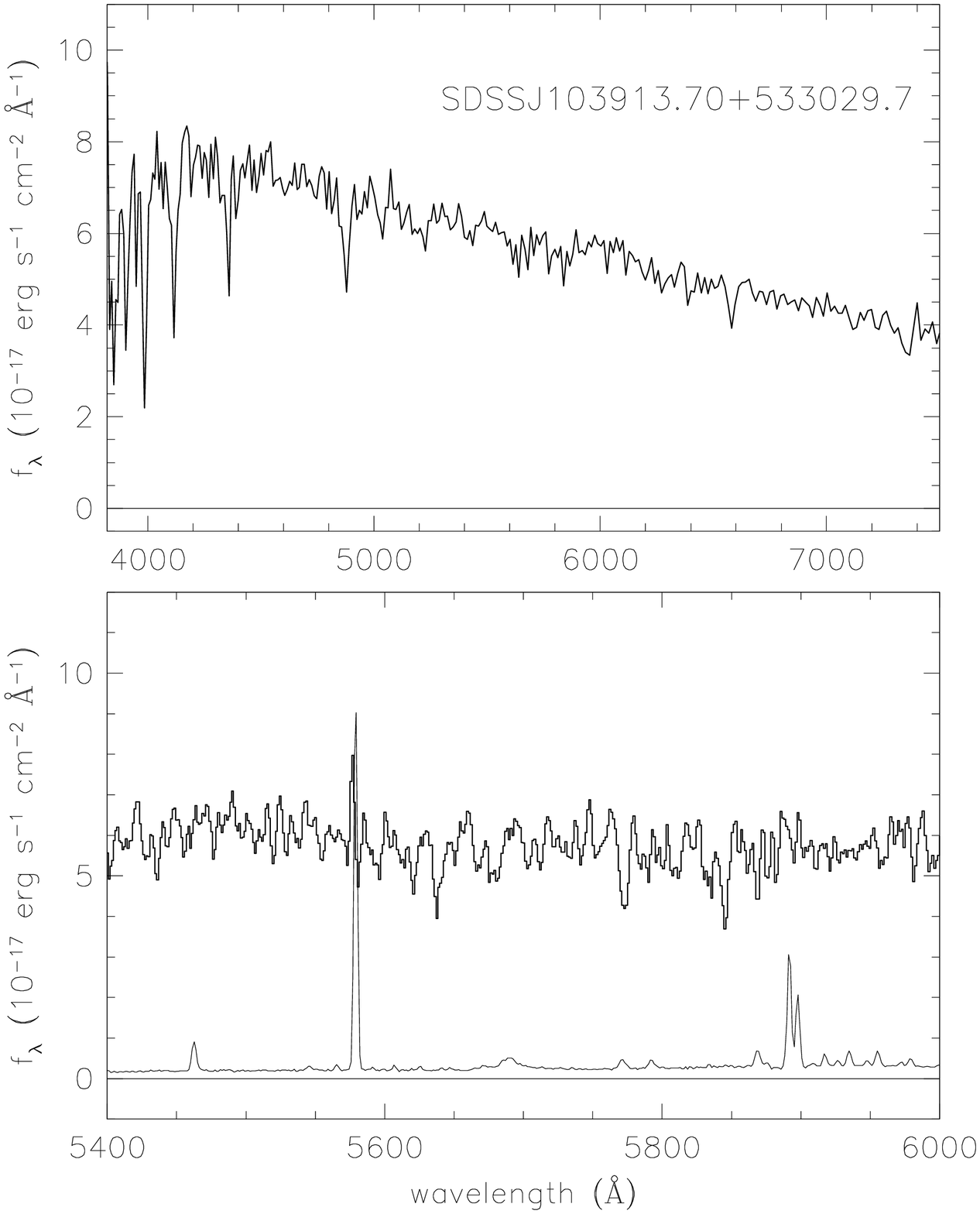}
%\includegraphics[angle=270]{spectrum.ps}
%\dfplot{f1.eps}
\figcaption{SDSS spectrum of SDSS1039+53; the spectral resolution
is $\approx$ 2000.
Lower panel: expanded region of the spectrum
near the atmospheric HgI 5460, [OI] 5577 and NaI D lines.
The light line shows the sky spectrum, where these lines are observed
at the rest wavelengths.
The spectrum is from plate 1010, fiber 335, MJD 52649, and is available
at the SDSS DR3 website (http://www.sdss.org).
\label{fig1}
}
\end{figure}
%----------------------------------------------------------------------
%
%----------------------------------------------------------------------
%  FIGURE 2  Color color plot
%----------------------------------------------------------------------
\begin{figure}[tb]
\figurenum{2}
\epsscale{0.9}
\plotone{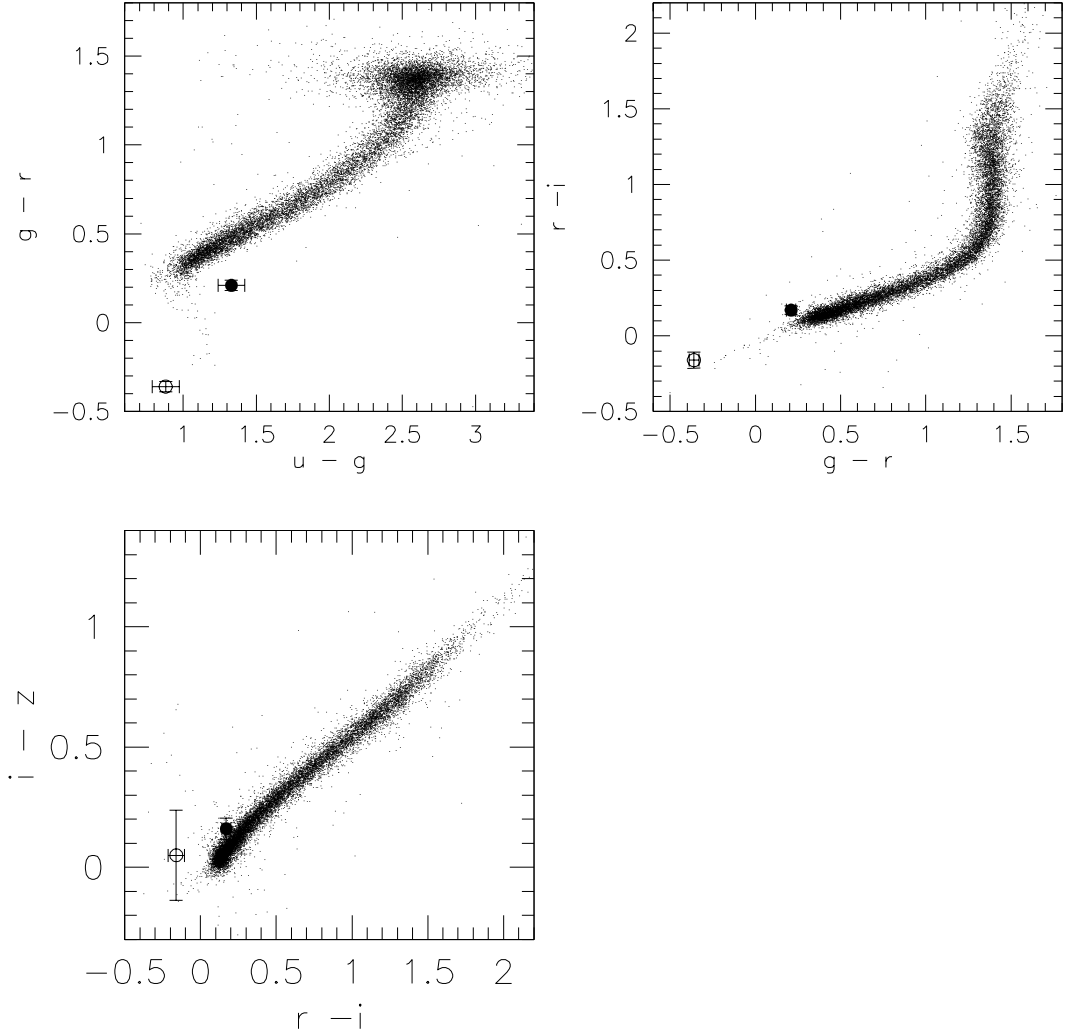}
%\dfplot{f2.eps}
\figcaption{SDSS color-color plots. The dots show the colors of a 
sample of $\sim$ 15,000 anonymous stars measured by SDSS (see 
Finlator et al. 2000). The large filled symbol shows the colors of the 
super star cluster SDSS1039+53 and the open symbol the colors of 
SDSSJ090745.0+024507, the hyper-velocity Galactic star described by 
Brown et al. (2005) (see text for discussion). Note that the colors of
SDSS1039+53 are similar, but not identical, to those of F stars.
\label{fig2}
}
\end{figure}
%----------------------------------------------------------------------
%
%----------------------------------------------------------------------
%  FIGURE 3  Image
%----------------------------------------------------------------------
\begin{figure}[tb]
\figurenum{3}
\epsscale{1.1}
\plotone{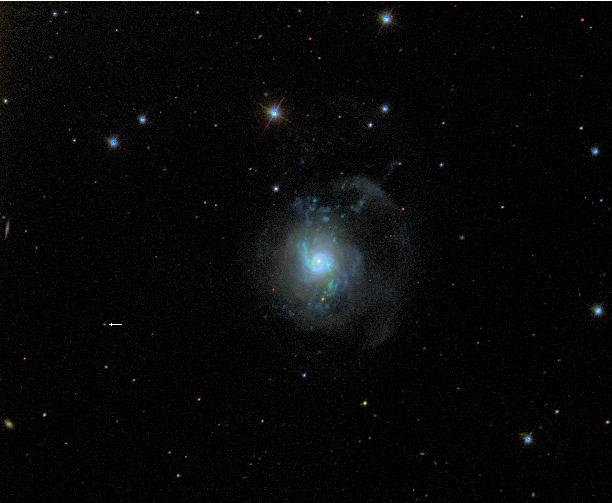}
\figcaption{Color composite image of NGC 3310 and its environment from
SDSS imaging.  The image is a composite of the SDSS $g$, $r$ and $i$
images made using the color weighting scheme described by Lupton et al.
(2004) and measures $13.6' \times 9.2'$. Roughly, north is up
and east to the left: the image is tilted at a position angle of about
$-20^{\circ}$.  The super star cluster SDSS1039+53 is to the left of the 
galaxy and is indicated by an arrow. The faint red streak running down the left
of the image is due to a 5th magnitude star to the north of the field.
The color mapping is such that H$\alpha$ is green and H$\beta$
+ O[III]$\lambda$5007 is blue, showing the presence of star forming regions.
Note the disturbed spiral arms and the inner ring of intense star formation.
\label{fig3}
}
\end{figure}
%----------------------------------------------------------------------
%
%----------------------------------------------------------------------
%  FIGURE 4  MODEL FITS
%----------------------------------------------------------------------
\begin{figure}[tb]
\figurenum{4}
\epsscale{0.7}
\plotone{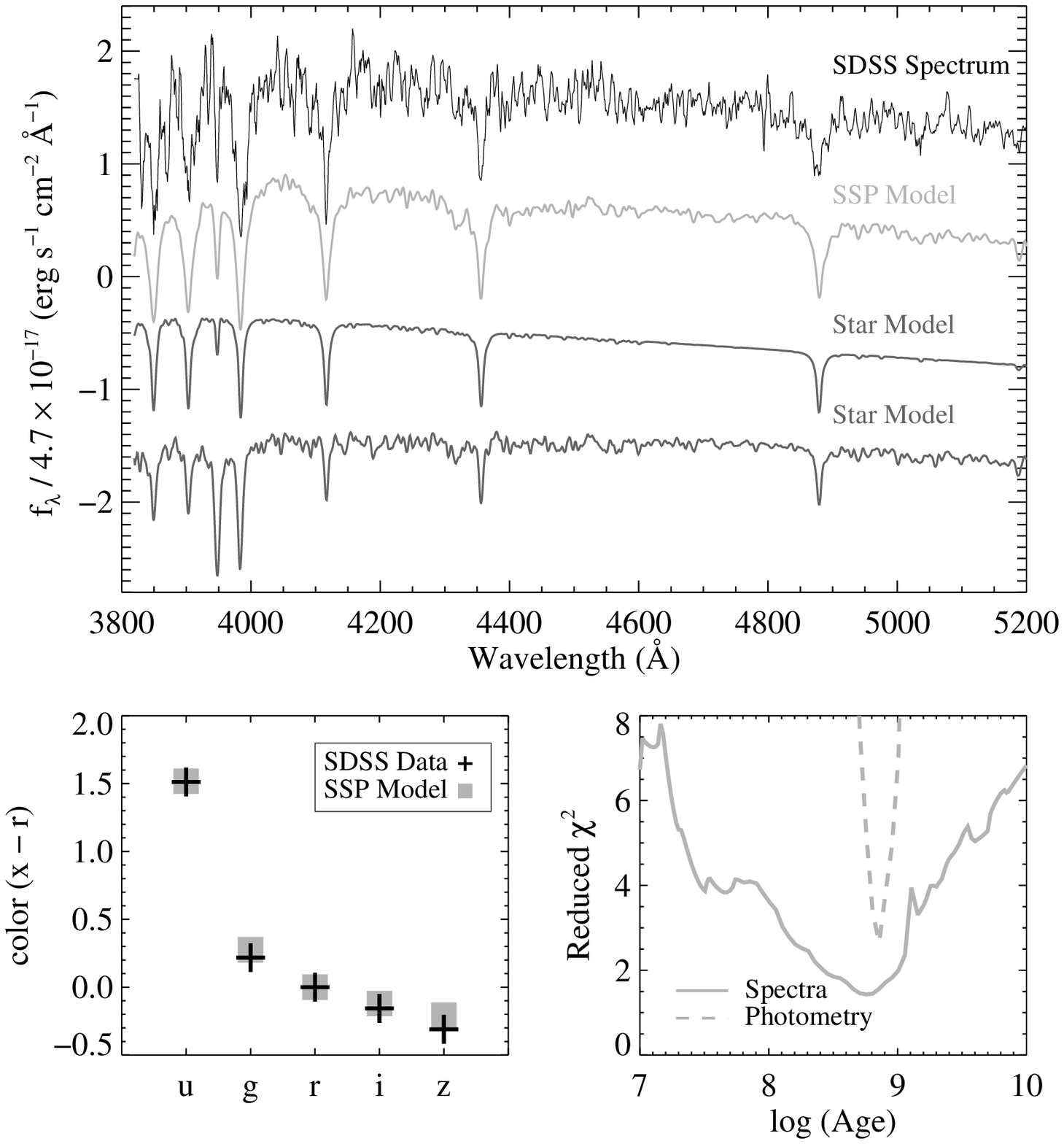}
%\includegraphics[angle=90,scale=0.70]{christy.ps}
%\dfplot{f4.eps}
\figcaption{Model fits to SDSS1039+53.  The upper panel contrasts the SDSS 
spectrum with the best fitting Simple Stellar Population (SSP) model, and 
two stellar models.  The SSP model is from Bruzual \& Charlot (2003)  and 
has an age of 570 Myr and Z = 1/4 Z$_{\odot}$.  The stellar models are from 
Kurucz (1993)
and have T$_{eff}$ = 6500 K, [Fe/H] = -2.5, log(g)  = 2.25 (upper), 
and T$_{eff}$ = 6250 K, [Fe/H] = -0.5, log(g) = -2.0 (lower). Both the
data and the models have been boxcar smoothed by 5 pixels.  The SSP model 
provides a considerably better fit to the the Ca~K line at 3934~\AA.  The 
lower left panel compares the SDSS photometry with the best fitting SSP 
model.  The error bars on the colors of SDSS1039+53 are smaller than the 
data points. The lower left panel shows the evolution of the reduced 
$\chi^2$ of the fits to both the SSP spectrum (solid) and the SSP 
photometric data (dashed).  The two independent fits provide some feel for 
the systematic errors.  We estimate the cluster age to be 700$\pm$150 
Myr.
\label{fig4}
}
\end{figure}
%----------------------------------------------------------------------
%
%------------------------------------------------------------------------------
% ACKNOWLEDGMENTS
%------------------------------------------------------------------------------

We thank the referee for a prompt and helpful report.
Partial support for the computer systems required to process and store 
the data was provided by NASA via grant NAG5-6734
and by Princeton University. 
We also thank Princeton University and the National Science Foundation
via grant AST03-07582 for generous support.
T.C.B. acknowledges partial support of this work from grants AST 
04-06784, and PHY 02-16783, Physics Frontier Centers/JINA: Joint 
Institute for Nuclear Astrophysics, awarded by the US
National Science Foundation.
This research made use of the IDL Astronomy User's Library at Goddard.

Funding for the creation and distribution of the SDSS Archive has been 
provided by the Alfred P. Sloan Foundation, the Participating 
Institutions, the National Aeronautics and Space Administration, the
National Science Foundation, the U.S. Department of Energy, the
Japanese Monbukagakusho, and the Max Planck Society. The SDSS Web site
is http://www.sdss.org/. The SDSS is managed by the Astrophysical
Research Consortium (ARC) for the Participating Institutions. The
Participating Institutions are The University of Chicago, Fermilab, 
the Institute for Advanced Study, the Japan Participation Group,
The Johns Hopkins University, the Korean Scientist Group, Los Alamos
National Laboratory, the Max-Planck-Institute for Astronomy (MPIA), 
the Max-Planck-Institute for Astrophysics (MPA), New Mexico State 
University, University of Pittsburgh, University of Portsmouth,
Princeton University, the United States Naval Observatory, and 
the University of Washington.

The Hobby-Eberly Telescope (HET) is a joint project of the University of
Texas at Austin, The Pennsylvania State University,  Stanford University,
Ludwig-Maximillians --Universit\"at M\"unchen, and Georg-August-Universit\"at
G\"ottingen.  The HET is named in honor of its principal benefactors,
William P. Hobby and Robert E. Eberly.  The Marcario Low-Resolution
Spectrograph is named for Mike Marcario of High Lonesome Optics, who
fabricated several optics for the instrument but died before its completion;
it is a joint project of the Hobby-Eberly Telescope partnership and the
Instituto de Astronom\'{\i}a de la Universidad Nacional Aut\'onoma de M\'exico.

\newpage
%------------------------------------------------------------------------------
% REFERENCES
%------------------------------------------------------------------------------

\bibliographystyle{unsrt}
\bibliography{gsrp}

%------------------------------------------------------------------------------
% TABLES
%------------------------------------------------------------------------------

\clearpage
%------------------------------------------------------------------------------
\begin{deluxetable}{rrrrrrrrrr}
\footnotesize
\tablewidth{0pt}
\tablecaption{SDSSJ103913.70+533029.7
   \label{table_conversions}
}
\tablehead{
\colhead{} &
\colhead{} &
\\
\colhead{} &
\colhead{} &
}
\startdata
   $\alpha$(J2000)& $\rm 10^h~39^m~13.70^s$& \\
   $\delta$(J2000)& $\rm +53^{\circ}~30'~29.7''$ \\
   $\rm v_{helio}$& 1012 $\pm$ 19 $\rm km~s^{-1}$ \\
   $\rm v_{GSR}$& 1067 $\pm$ 19 $\rm km~s^{-1}$ \\
   $u$& 20.90 $\pm$ 0.09\\
   $g$& 19.57 $\pm$ 0.02 \\
   $r$& 19.36 $\pm$ 0.02\\
   $i$& 19.19 $\pm$ 0.02\\
   $z$& 19.03 $\pm$ 0.04\\
\enddata
\tablecomments{The magnitudes are corrected for interstellar extinction
using the data of Schlegel, Finkbeiner \& Davis (1998).}
\end{deluxetable}
\clearpage
\end{document}